\documentclass{article}

\usepackage{graphicx}
\usepackage{psfig}
\usepackage{epsfig}
\usepackage[round]{natbib}

\title{A New Framework for the Assessment and Calibration of Medium Range Ensemble Temperature Forecasts}

\author{Stephen Jewson, Anders Brix and Christine Ziehmann\footnote{\emph{Correspondence address}: RMS, 10 Eastcheap,
London, EC3M 1AJ, UK. Email: \texttt{x@stephenjewson.com}}\\
London, United Kingdom}

\begin{document}

\maketitle

\begin{abstract}

We present a new framework for the assessment and calibration of medium range ensemble temperature forecasts.
The method is based on maximising the likelihood of a simple parametric model for the temperature distribution,
and leads to some new insights into the predictability of uncertainty.

\end{abstract}

\section{Introduction}

A number of different methods have been used for the assessment
and calibration of ensemble forecasts (for example
see~\citet{atger99}, \citet{richardson00}, \cite{roulstons02} and
\cite{wilsonet98}). In many applications of ensemble forecasts the
forecast is used to derive the probability of a certain outcome,
such as temperature dropping below zero. In this context, the
reliability diagram is an appropriate method for assessing
reliability (see~\cite{anderson96}, \cite{eckelw98},
\cite{talagrandet97} and~\citet{hamill97}) and the relative
operating characteristic (ROC) (\cite{mason82}, \cite{swets88},
\cite{masong99}) is an appropriate method to evaluate the
resolution.

In other applications of ensemble forecasts, however, the forecast
is interpreted as providing a mean and a distribution of future
values of temperature. For example in the field of weather
derivatives the calculation of the fair strike for a certain class
of weather swap contract\footnote{linear swaps on a linear
temperature index} needs an estimate of the mean of the future
temperatures, while the calculation of the fair premium for
weather option contracts needs an estimate of the whole
distribution of future temperatures (see~\citet{jewsonc02b} for
details). Additionally, the assumption is often made that
temperature is normally distributed since this allows the
temperature forecast to be summarised succinctly using just the
mean and the standard deviation. For such mean-and-distribution or
mean-and-standard deviation based applications of ensemble
forecasts the reliability diagram and the ROC are not particularly
appropriate.

In this paper we present a new framework for the assessment and
calibration of ensemble temperature forecasts based on analysis of
the mean and standard deviation of the distribution of
temperatures. The method has been developed to respond to the need
for a simple and practical method for assessment and calibration
that can be used by companies that make use of ensemble forecasts
in the weather derivative market. We postulate a parametric model
for the mean and standard deviation and fit the parameters of the
model using the maximum likelihood method. This approach has a
number of advantages relative to the assessment and calibration
methods mentioned above. The model is simple, easy to interpret,
and the entire ensemble distribution can be calibrated in one
simple step. Also the model gives a clear indication of how many
days of useful information there are in a forecast.

In section 2 we describe the data sets we use for this study. In
section 3 we describe the statistical model that forms the basis
for the method we propose. In section 4 we describe the results
from fitting the model. In section 5 we discuss extensions to
other distributions and in section 6 we summarise our results and
draw some conclusions.

\section{Data}

We will base our analyses on one year of ensemble forecast data for the weather
station at London's Heathrow airport, WMO number 03772. The forecasts are predictions
of the daily average temperature, and the target days of the forecasts
run from 1st January 2002 to 31st December 2002. The forecast was produced
from the ECMWF model~\citep{molteniet96} and downscaled to the airport location using a simple
interpolation routine prior to our analysis. There are 51 members in the ensemble.
We will compare these forecasts to the quality controlled climate
values of daily average temperature for the same location as reported by the UKMO.

Throughout this paper all equations and all values have had both the seasonal mean and
the seasonal standard deviation removed. Removing the seasonal standard deviation
removes most of the seasonality in the forecast error statistics, and justifies the use of
non-seasonal parameters in the statistical models for temperature that we propose.

\section{The Moment-based Ensemble Assessment and Calibration Model}

For forecasts of temperature anomalies, it has long been recognized
(see for example~\citet{Leith})
that the use of a final regression step between ensemble mean and observations
can eliminate bias and minimise the mean square error (MSE).
For our purposes we will write this regression step as:

\begin{equation}\label{regression}
 T_i \sim N(\alpha+\beta m_i, \sigma)
\end{equation}

where $T_i$ is the observed temperature on day $i$,
$N(\mu,\sigma)$ represents a normal distribution with mean $\mu$
and standard deviation $\sigma$, $m_i$ is the forecast of the
temperature (in our case, the ensemble mean) and $\alpha$, $\beta$
and $\sigma$ are free parameters. This regression model postulates
that temperatures come from a normal distribution with mean given
by $\mu_i=\alpha+\beta m_i$ and standard deviation given by
$\sigma$. The values for $\alpha$, $\beta$ and $\sigma$ come from
fitting the model, and this is usually done using least squares
linear regression. One justification for the use of least squares
linear regression is that for this particular model it is
equivalent to finding the parameters that maximize the likelihood
of the data given the model (see~\citet{nr}), as long as we assume
that the forecast errors are uncorrelated in time. We note that
although the model in equation 1 postulates that the data come
from a normal distribution, it can be applied in situations in
which the data is not strictly normal, and in fact it is common
(although perhaps bad) practice not to test for normality when
doing such linear regressions.

One of the assumptions in this model is that the standard
deviation of the forecast errors $\sigma$ is constant. However it
is well documented that the size of forecast errors varies in
time~\citep{palmert88} and that there is a relationship between
the ensemble spread and the size of forecast errors~\citep{totht00}.
It thus makes sense to attempt to generalize the model in
equation~\ref{regression} to a model that takes these temporal
variations in $\sigma$ into account. We will do this using the
model:

\begin{equation}\label{model}
 T_i \sim N(\alpha+\beta m_i, \gamma+\delta s_i)
\end{equation}

where the free parameter $\sigma$ has been replaced by a linear function
of the ensemble spread $s_i$, and two new parameters
$\gamma$ and $\delta$ have been introduced.
Modelling the standard deviation
as a linear function of the ensemble spread in this way allows for both time variation and
the correction of biases in the predicted uncertainty.\footnote{We note that one could
alternatively model the variance as a linear function of the spread squared.}

The optimum parameters for this model can no longer be fitted
using least squares linear regression. However, they can be fitted
if we can identify a cost function that can be minimised or
maximised by varying the parameters. There are various
possibilities for such a cost function, but one of the most
natural is the likelihood, defined as the probability density of
the observations given the calibrated forecast. Maximising the
likelihood is the standard way to fit parameters in statistics
(see for instance textbooks such as~\citet{casellab02}
or~\cite{lehmannc98}), and gives the most accurate possible
estimates of the parameters for most statistical models.

As with the linear regression model, this model is also not
restricted to cases in which temperature is exactly normally
distributed: the assumption of the normal distribution merely
provides a metric in which the likelihood can be calculated and
the parameters fitted. This metric is most appropriate when the
data is at least close to normally distributed. For cases when the
data is not close to normal other distributions can be used, or
the data can be transformed to normal.

There are a number of useful features of the model we present. These include:

\begin{itemize}
  \item
Once the parameters have been fitted to past historical data, calibration
of future ensemble forecasts is easy since it just involves
applying linear transformations to the ensemble mean and standard deviation.
The calibrated values for the mean and the standard deviation
can be used to define the whole forecast distribution,
or can be used to shift and stretch the individual ensemble members,
if individual ensemble members need to be preserved.
In the latter case non-normality in the distribution of the
original ensemble members will not be destroyed.
  \item
The optimum values of the parameters in equation 2 have clear
interpretation and give us useful information about the
performance of the ensemble. For instance $\alpha$ identifies a
bias in the mean, and $\beta$ represents a scaling of the forecast
towards climatological values. In a perfect forecast, $\alpha$
would be zero and $\beta$ would be one. The spread parameters
$\gamma$ and $\delta$ combine to optimize the prediction of
uncertainty about the mean. The value of the ensemble spread $s$
varies in time because of the dependence of the growth rate of
differences between ensemble members on the actual model state.
The calibrated standard deviation value $\sigma_i=\gamma+\delta
s_i$ additionally includes uncertainty due to model error. If the
spread of the ensemble contains very little real information,
$\delta$ will tend to be small, and $\gamma$ will tend to be large
to compensate.

  \item
It is very easy to calculate approximate uncertainty levels on the
values of the parameters as part of the fitting procedure. This is
done using the curvature of the log-likelihood at the maximum (see
the above references on likelihood methods). These uncertainty
levels give us a clear answer to the question of whether the
ensemble forecast has useful skill at different lead times. For
instance, once $\beta$ is not significantly different from zero we
can say that the ensemble mean no longer contains useful
information (at least not within this framework) and once $\delta$
is not significantly different from zero then we can say that the
ensemble spread no longer contains useful information. This raises
the interesting possibility that we might identify situations in
which the mean may contain more days of useful information than
the spread.

 \item It is often necessary to decide which of two forecasts is
 the more accurate. If two forecasts are both calibrated using
 equation~\ref{model} then the log-likelihood provides a natural way
 to compare the forecasts. Log-likelihood measures the
 ability of the forecast to represent the whole distribution of
 observed temperatures, and is a generalisation of mean square error.
 It can be presented in a number of ways
 such as log-likelihood or log-likelihood skill score.
\end{itemize}

Forecasts calibrated using equation~\ref{model} will not necessarily minimise MSE. Users
interested purely in a single forecast that minimises MSE should thus calibrate using
equation~\ref{regression}. However, users interested in predictions of uncertainty, or,
equivalently, in the whole distribution of possible temperatures, should calibrate using
equation~\ref{model}. In practice we have found that the mean temperature prediction
produced by equation~\ref{model} is close to that produced by equation~\ref{regression},
presumably because the fluctuations in uncertainty are not large.

\section{Results}
\label{results}

\begin{figure}[!htb]
  \begin{center}
    \scalebox{0.9}{\includegraphics{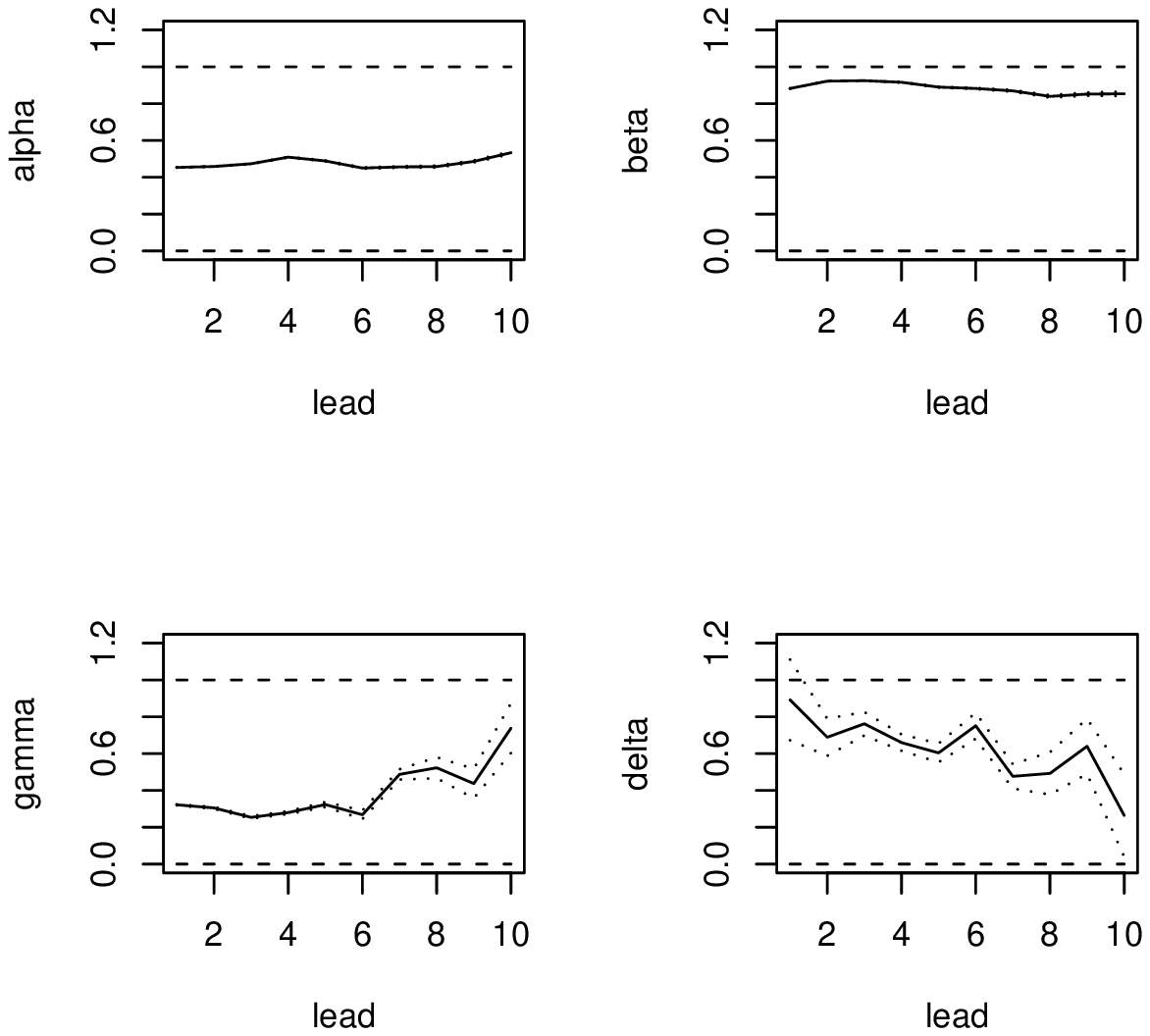}}
  \end{center}
 \caption{The optimum values for the parameters in equation~\ref{model} (solid line),
 95\% confidence intervals (dotted line) and the constant values 0 and 1 (dashed line)}
 \label{f:params}
\end{figure}

The optimum values for the parameters in equation 2 for our 1 year
of forecast data and observations are shown in
figure~\ref{f:params}. In each case we show the approximate 95\%
sampling error confidence intervals around the optimum parameters.
In some cases they are so narrow that they are hard to see in the
graphs.

Looking at $\alpha$ we see that there is a small and roughly constant bias in the
temperatures produced by the ensemble. Correction of the ensemble mean (or each ensemble
member) using $\alpha$ would eliminate this bias, as long as the ensemble stays stationary.

The parameter $\beta$ is slightly less than 1 at all leads. This
shows that the ensemble mean varies too much: either the ensemble
mean, or each ensemble member, should be reduced by the factor
$\beta$ towards the climatology. Such a damping factor is
presumably required because the ensemble members are more
correlated with each other than they are with the observations and
because the ensemble is finite in size. Even at lead 10 $\beta$ is
highly significantly different from zero, implying that the
ensemble mean still contains useful predictive ability at that
lead. If we allow ourselves to extrapolate the $\beta$ curve to
longer leads by eye, it would seem likely that the ensemble mean
would still contain useful predictive information even beyond
that.

The fact that our values of $\delta$ are significantly different from
zero out to the end of the forecast (just) shows that there is significant information in the ensemble spread too.
However, in this case if we extrapolate to higher lead times by eye it seems unlikely that
there would be any more skill in $\delta$.
Since $\delta$ is below one and $\gamma$ is non-zero
we see that the standard deviation of the ensemble
is not an optimal estimate of the uncertainty of the prediction.

\begin{figure}[!htb]
  \begin{center}
    \scalebox{0.9}{\includegraphics{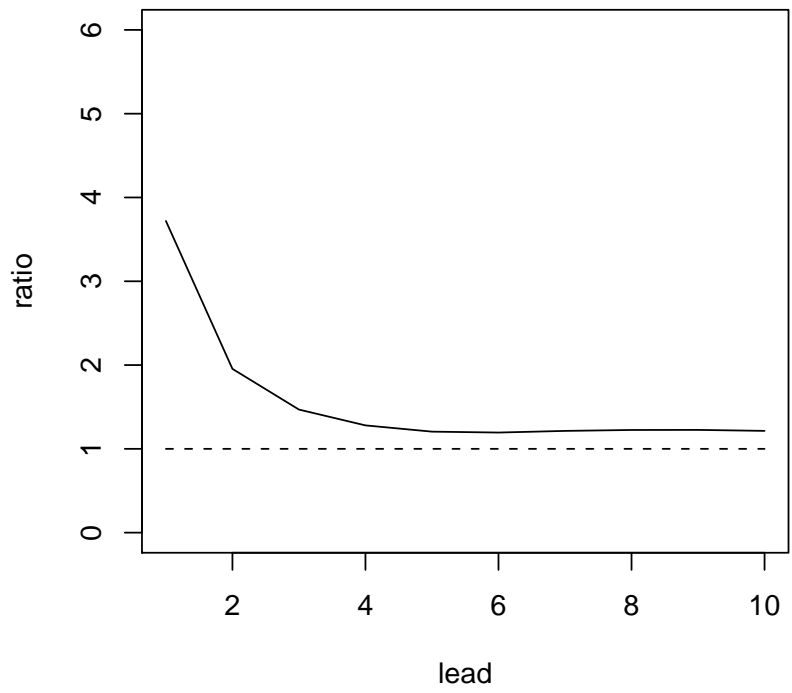}}
  \end{center}
 \caption{The ratio of the time mean of the standard deviation of the calibrated ensemble to that of the
 uncalibrated ensemble.}
 \label{f:ratio}
\end{figure}

The $\gamma+\delta s$ transformation can change both the mean
spread (the time mean of the standard deviation across the
ensemble) and the variability of that spread (the standard
deviation in time of the standard deviation across the ensemble).
To measure the effect on the mean spread, figure~\ref{f:ratio}
shows values of $\frac{\gamma+\delta \overline{s}}{\overline{s}}$
(where the overbar indicates the mean in time over the year of
data) which shows the factor by which the transformation increases
the mean spread. We see that at short lead times, the ensemble
spread $s$ is far too small on average and the calibration
increases the spread by factors of around 4 (at lead 0) and 2 (at
lead 1). At longer lead times the ensemble spread is still too
small on average by a factor of around 1.2. This underestimation
of the spread from ensemble forecasts has been noted by a number
of authors such as~\citet{ziehmann00} and~\citet{mullenb}. It is
likely to be due to model error in the prediction model and due to
the fact that the forecast is a prediction of a large scale flow
while the observation is site-specific and hence affected by
small-scale variability not represented in the model.

The size of the effect of the calibration on the \emph{variability} (in time) of the spread
is given by the value of $\delta$.
Since $\delta$ is significantly different from one at all lead times beyond the first
we conclude that the variability of the spread
from the ensemble needs to be reduced to be optimum at those lead times.
This could be because the variability
of the ensemble spread is too large, or because the variability of the ensemble spread
is not highly correlated with the real variability of skill.

We can see from the values of $\delta$ that the variability
of the ensemble spread alone will overestimate the state-dependent predictability of this model by a large
factor at long leads.
A better estimate for the level of state-dependent predictability is given by the variability of the calibrated
spread, which is smaller by the $\delta$ factor.

\begin{figure}[!htb]
  \begin{center}
    \scalebox{0.9}{\includegraphics{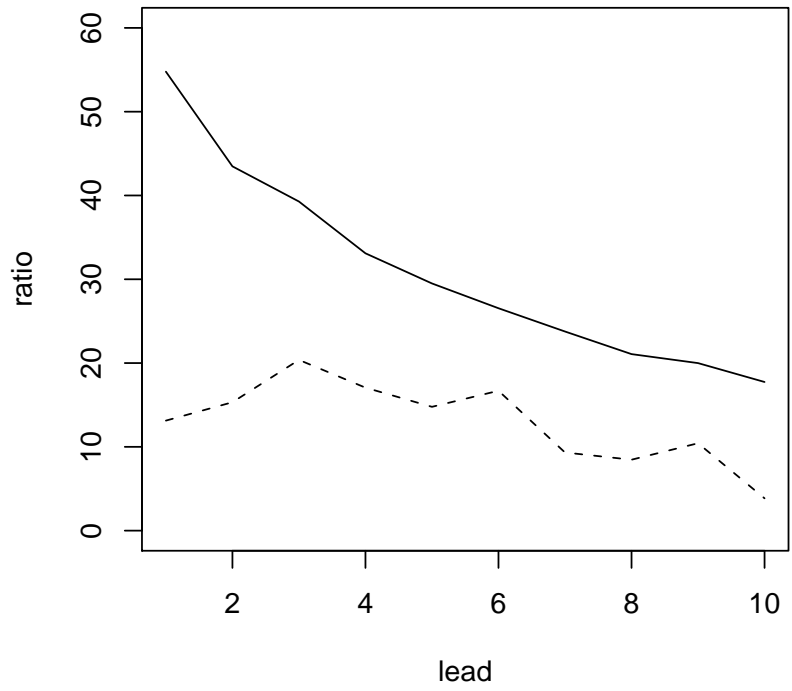}}
  \end{center}
 \caption{Both lines show the ratio of the standard deviation in time of the standard deviation across the ensemble
 to the mean in time of the standard deviation of the ensemble. This ratio is given the name coefficient of variation
 of spread (COVS) in the text.
 The solid line was estimated using the uncalibrated ensemble, and the dotted line using the calibrated
 ensemble.}
 \label{f:ratio2}
\end{figure}

Figure~\ref{f:ratio2} shows the ratio of the standard deviation of the ensemble spread to
the mean ensemble spread at different lead times. We call this ratio the coefficient
of variation of the spread (COVS). Figure~\ref{f:ratio2} shows the COVS estimated from both the
uncalibrated and the calibrated ensemble
data. These values give an indication of how much extra information we get about the forecast uncertainty
by using the (uncalibrated or calibrated) spread of the
ensemble rather than using a level of uncertainty which is constant with time.
The uncalibrated data suggests that variations in uncertainty that are 20\% to 55\% of the mean uncertainty are
predictable using the ensemble spread.
However, because the uncalibrated data both underestimates the total spread
(the numerator in the COVS)
and overestimates the
predictable part of the variability of the spread (the denominator in the COVS)
these values seem to be overestimates. The calibrated data suggests
that variations in the uncertainty that are only 5\% to 20\% of the mean uncertainty are
predictable using the ensemble spread.

\section{Other distributions}
\label{futurework}

In cases where the forecast errors are not close to normally
distributed, one can use other distributions. For example in the
case where the forecast errors show skew the skew-normal
distribution $SN$ can be used. The skew-normal distribution is a
generalisation of the normal distribution which has a third
parameter, and includes the possibility of modelling
skew~\citep{azzalini85}. Suppressing the index $i$ for clarity we
then have:

\begin{equation}\label{sn}
 T \sim SN( \alpha+\beta m, \gamma+\delta s, \zeta+\eta k)
\end{equation}

where we have introduced the ensemble skew $k$ and two new parameters $\zeta$ and $\eta$.

The skew-normal model
can be fitted using maximum likelihood methods exactly as for the normal distribution.
One of the
results from such a fitting process would be a clear indication as to whether the
forecast being calibrated does or does not contain statistically significant
information about the skew of observed temperatures (this question has been discussed
by~\citet{mylne02b}).

For extremely non-normal distributions for which even the skew-normal is not non-normal enough,
non-parametric distributions may be more appropriate. A simple non-parametric method
would be to use a kernel density, with a single free parameter for the width of
each kernel (see~\citet{bowmana} for a description of kernel densities).
Such a method would look a little like the method of~\citet{roulstons03} even though
it is justified in a completely different way.

\section{Conclusions}

We have described a simple parametric method for the assessment
and calibration of ensemble temperature forecasts. The method
consists of applying linear transformations to the mean and
standard deviations from the ensemble. The parameters of the model
can be fitted easily using the maximum likelihood method. The
model has various advantages and disadvantages relative to other
calibration models currently in use. The main disadvantage is that
the model only works for forecast errors that are reasonably close
to normally distributed, although extensions have been described
that should overcome that limitation. The advantages of the model
are that:

\begin{itemize}
  \item the calibration of forecasts using the model is extremely simple
  \item the model is transparent and easy to understand
  \item the model separates skill in predictions of the mean and the spread
  \item calculating confidence intervals on parameters is easy
  \item the model gives a clear indication of how many days of useful skill there are in a forecast
\end{itemize}

We have applied the model to one year of site-specific ECMWF ensemble forecasts.
We find that the forecasts have highly significant skill for predicting both the mean and the
standard deviation out to 10 days. The
forecasts underestimate the mean uncertainty, as has been reported in other studies.
They also over-estimate the variability of the uncertainty.
For these forecasts we estimate that the predictable part of the uncertainty is only
between 5\% and 20\% of the mean uncertainty, depending on lead time. For some applications
this variability in the uncertainty may be small enough that it can be ignored and one
could make the simplifying assumption that the uncertainty is constant in time.

Further work includes:

\begin{itemize}

  \item Developing algorithms that avoid having to make the assumption that
        the forecast error is uncorrelated in time.

  \item Out of sample testing of the calibrated forecasts, using both measures from within the
        framework (i.e. likelihood) and also other measures such as rank histograms, reliability diagrams and
        ROCs.

\end{itemize}

\section{Acknowledgements}

We would like to thank K. Mylne for providing us with the forecast data, Risk Management Solutions for
providing us with the observational data, and S. Mason, F. Doblas-Reyes and D. Anderson for useful discussions
on the subject of ensemble validation. This research was funded by the authors.

\bibliography{jewson}

\clearpage

\clearpage

\clearpage

\end{document}